\def\spose#1{\hbox to 0pt{#1\hss}} \def\lta{\mathrel{\spose{\lower
3pt\hbox{$\mathchar"218$}} \raise 2.0pt\hbox{$\mathchar"13C$}}}
\def\gta{\mathrel{\spose{\lower 3pt\hbox{$\mathchar"218$}} \raise
2.0pt\hbox{$\mathchar"13E$}}}
 \def\etal{{\it et al.\ }}
\def\frac#1#2{\leavevmode\kern.1em \raise.5ex\hbox{\the\scriptfont0
  #1}\kern-.1em /\kern-.15em\lower.25ex\hbox{\the\scriptfont0 #2}}
   \def\third{\frac{1}{3}}
   \def\twothirds{\frac{2}{3}}
 
\def\deg{^\circ}             %for angular measure in degrees

  \def\msun{{\rm M}_\odot}
\def\lsun{{\rm\,L_\odot}}  \def\yr{{\rm\,yr}}

\def\hb{\hfill\break}
 
\documentstyle[11pt,paspconf]{article}

\begin{document}

\title{Interacting and Merging Galaxies}

\author{Uta Fritze - von Alvensleben} \affil{Universit\"atssternwarte
G\"ottingen, Geismarlandstra{\ss}e 11, 37083 G\"ottingen, Germany}

% Notice that some of these authors have alternate affiliations, which
% are identified by the \altaffilmark after each name.  The actual alternate
% affiliation information is typeset in footnotes at the bottom of the
% first page, and the text itself is specified in \altaffiltext commands.
% There is a separate \altaffiltext for each alternate affiliation
% indicated above.

% The abstract is entered in a LaTeX "environment", designated with paired
% \begin{abstract} -- \end{abstract} commands.  Other environments are
% identified by the name in the curly braces.

% Poster authors ONLY may omit the abstract in order to gain a little
% more page space for the text of the poster.

\begin{abstract}
I will present a review of some of the many recent developments within
the exciting and rapidly evolving field of Interacting and Merging
Galaxies. I will touch both on observations and on theory, focussing
on those aspects where galaxy evolution modelling may possibly
contribute to our understanding.  \hb After briefly outlining the
basic concepts and a few specific results of stellar- and gasdynamical
modelling of galaxy -- galaxy mergers (Sect. 1) I choose three
examples of interacting/merged galaxy pairs to discuss different
aspects of mergers: tracing back the star formation history of the old
merger remnant NGC 7252 (Sect. 2), the formation of young star
clusters (Sect. 3) and dwarf galaxies (Sect. 4) in ``The Antennae'',
and the molecular gas content of IR-UL galaxies on the example of Arp
220 (Sect. 5).  I'll touch the interaction -- activity connection in
Sect. 6 and report attempts to identify merger remnants among today's
galaxies in Sect. 7. After few words on merger rates at present and in
the past (Sect. 8), I'll close with a brief outlook.

\end{abstract}

% Keywords should be included, but they are not printed in the hardcopy.


\section{How it all began... }
Arp's Atlas of Peculiar Galaxies (1966) shows an impressive variety of
disturbed galaxies, many of which are evidently interacting or
merging. The seminal work of Toomre \& Toomre (1972) put forward a
whole bunch of ideas which keep stimulating research since more than
20 years. Their basic idea was that the Hubble sequence of galaxies
need not be given {\it ab initio ad infinitum}, i.e. that Hubble types
may change in the course of strong interactions. With a restricted
3-body algorithm for an N$\sim120\,-\,$body system of particles they
modelled the merging of two spirals and showed that the remnant may
look very much like an elliptical.  Comparing their models to nearby
examples of apparently interacting/merging galaxies -- most from Arp's
atlas -- they established their famous {\it age sequence of the
eleven}. The earliest stage of interaction is represented by Arp 244 =
NGC 4038/39 = ``The Antennae'' and the latest stage of a completed
merger by Arp 226 = NGC 7252, both of which I'll present in some
detail in Sect. 2 \& 3. Estimating the galaxy merger rate among NGC
galaxies they found the number of galaxies that should have
experienced a major merger over a Hubble time to be comparable to the
number of E/S0 galaxies, thus raising the suspicion that some or even
all E/S0 galaxies might be produced by Sp -- Sp mergers. Furthermore,
they already considered a possible interaction -- activity connection,
``{\it stoking the furnace}'', which I will mention in Sect. 6. \hb
From the very beginning, the idea that ellipticals may be built by Sp
-- Sp mergers has provoked vigorous {\bf C}ounter-{\bf A}rguments, the
most famous ones being
\begin{description}
\item[CA 1:] Central densities in ellipticals are much higher than in
spirals,
\item[CA 2:] surface brightness profiles are distinctly different:
deVaucouleurs - like in E's vs exponential in spiral disks,
\item[CA 3:] the existence of gradients within and of a luminosity --
metallicity relation among ellipticals,
\item[CA 4:] the specific globular cluster ({\bf GC}) frequencies,
i.e. the number of GCs normalised to the stellar mass of the parent
galaxy $T_{GC} := N_{GC} / M_{\ast}$ (Zepf \& Ashman 1993) is roughly
twice as high in a typical elliptical as compared to a typical spiral.
\end{description}

\noindent
Basically, two major {\bf Stellardynamical Approaches} have been
developed to model galaxy -- galaxy interactions: expansion codes
calculating the potential from basis function expansions of the
density field (Aarseth 1967 ff, van Albada \& Gorkom 1977, ...)  and
hierarchical TREE codes (Appel 1985, Jernigan 1985, Barnes \& Hut
1986, ...) computing the potential from distant groups of particles
using low-order multipole expansions. TREE codes have the advantage of
being gridless and keeping the action-at-a-distance concept.  The
number of particles treatable in N-body codes is increasing
tremendously in recent years from 250 (White 1978/79, N-Body) over
$10^4$ (Barnes 1988, TREE), $10^5$ (Hernquist 1993, TREE) to more than
$10^6$ (GRAPE).

Out of the enormous amount of insight these stellardynamical models
have provided, I'll pick a few points that we will need later to
understand the observations I'll present.  The nicest tails appear in
slow direct encounters in an isolated environment, only disks of
comparable mass can produce a pair of equal-length 359ails, up to
$25-50$ \% in mass of each disk can get into the tails.
Selfgravitating condensations that resemble dwarf galaxies are
frequently observed along the tails.  Orbital decay efficiently stirs
up material in both disks and leads to rapid merging.  Violent
relaxation ({\bf VR}) is able to produce a de$\,$Vaucouleurs profile
and to make the remnant obey the $L-\sigma$ relation, i.e. lie in the
fundamental plane of elliptical galaxies.  VR thus cancels CA
2. Moreover, VR is incomplete allowing gradients to partly survive
(cancelling CA 3) and propagates outward, leading to fall-back of
material from the tails, only few percent escape.  DM halos have been
shown to significantly increase the merger cross-section and to
considerably accelerate core merging by efficiently soaking up angular
momentun.  But still if $(Bulge + Disk + Halo)$ are selfconsistently
included in stellar-dynamical models the central densities of the
remnants are much lower than those in ellipticals. \hb
\noindent
However, observations show enormous concentrations of molecular gas in
the centers of interacting galaxies ($\rightarrow$ Sect. 5) and huge
HI envelopes around them (e.g. Sancisi 1995).

{\bf Gasdynamical Models} were developed along three basic streams:
{\bf S}mooth -- {\bf P}article {\bf H}ydrodynamics (Lucy 1977,
Monaghan 1992, ...), a Lagrangian gridless method, made adaptive in
space and time by Hernquist \& Katz (1989) is able to treat large
density contrasts.  In the Sticky Particles method (Negroponte \&
White 1983, ...) gas particles can dissipate energy in inelastic
collisions.  Both methods basically confirm and considerably detail
the principle results from early (semi-)analytic stability analyses of
gaseous disks that perturbations can lead to global instabilities
bringing along global and/or nuclear starbursts and eventually create
or feed a central black hole (Byrd \etal 1986, Lin \etal 1988).

Just to pick out of the large number of results from hydrodynamical
models a few items relevant to the impact of merging on the evolution
of the gaseous and stellar components of galaxies I like to recall
that tidal forces in a slow passage are able to perturb gas and stars
over the entire disk(s), that shocks efficiently transfer momentum
between parcels of gas, and that tidal perturbations, especially in
retrograde encounters of gas-rich spirals, can drive a large fraction
of all the gas close to the center.  During the final collision
hydrodynamic forces lead to coalescence of gas blobs and may end up
producing {\bf one} massive central gas cloud containing $\gta 50$\%
of all disk gas (cancelling CA 1). For an excellent extensive review
on the dynamical aspects of galaxy interactions the reader is referred
to Barnes \& Hernquist (1992a).

It is clear that in view of these gas concentrations {\bf Star
Formation (SF)} comes into play. Compression of gas clouds by
large-scale shocks or a hot surrounding medium as well as
precipitation of cloud -- cloud collisions are evoked to convert a
large fraction of gas into stars on a timescale of order $10^8$ yr
(e.g. Larson 1987, ..., Olson \& Kwan 1990, ...). A serious problem
for any dynamical model attempting to include SF is that the basic
physical processes of SF and the functional dependence of the Star
Formation Rate ({\bf SFR}) on gas densities, temperatures, etc. are
still poorly understood, in particular for the violent SF regime often
encountered in interacting/merging galaxies.

It is clear that massive SF has a significant feedback on the ISM in
terms of material and energetics, e.g. in the form of superwinds
(Heckman \etal 1987, ...). That this feedback may lead to some kind of
self-regulation might tentatively be concluded from the similar UV
peak surface brightnesses Meurer \etal (1996) find in 7 out of 9
starbursts.

Now, let's have a closer look at some famous examples.

\section{NGC 7252: an advanced post-burst with surprising activity}
With an estimated dynamical age of $\sim 1$ Gyr, NGC 7252 (= Arp 226)
(Schweizer 1982ff) is the oldest merger remnant on Toomre \& Toomre's
list. It features two impressive tidal tails of $\sim 100$ h$^{-1}$
kpc.  2D VLA HI mapping (Hibbard \etal 1993) shows velocity maxima
$\sim$ halfway along the tails containing huge amounts of HI
(M(HI)$\,\sim\,2\,\cdot\,10^9\,$h$^{-2} \msun$), while the center of
the merger remnant is essentially free of HI but harbors a
counterrotating disk of ionised and molecular gas
(M(H$_2$)$\,\sim\,2\,\cdot\,10^9\,$h$^{-2} \msun$) (Dupraz \etal 1990,
Wang \etal 1992). These observations prompted a new attempt of
dynamical modelling where Hibbard \& Mihos (1995) use a
self-consistent N=65\,000\,-body TREE code describing the gas in two
merging (disk+halo) systems and find a prograde initially parabolic
encounter to best fit all the available data and featuring a delayed
(up to a few Gyrs) fall-back of most of the gas in the tails to
successively larger radii within the main body of NGC 7252. As in many
other merger remnants, the light distribution in NGC 7252 follows a
de$\,$Vaucouleurs law and -- despite its impressive merger signatures
-- NGC 7252 lies unsuspectedly within the fundamental plain of normal
elliptical galaxies (Lake \& Dressler 1986). Prominent A-star spectral
features (Balmer absorption lines) in the nuclear and integrated
spectra as well as slightly blue UBVR colors that are constant over
$\sim 20$ kpc point to a strong global starburst $\sim 1$ Gyr ago
(H$_0=75$ kms$^{-1}$Mpc$^{-1}$ is used throughout).

To study the spectral and chemical evolution of merging galaxies
including their starbursts we developed a simplified 1-zone model for
the merger of two identical spirals. The 4 parameters of this model
are the spectral type (= the star formation history (SFH)) and age of
the progenitor spirals and the strength and duration of the tidally
triggered starburst. SFHs for undisturbed spiral types Sa, Sb, Sc, Sd
were shown to give good agreement after a Hubble time of the model
spectra with observed templates from Kennicutt's (1992) atlas. The 4
parameters were studied in terms of 2-color diagrams $UBVRIJHKL$ in
Fritze -- v. A. \& Gerhard (1994a). We use this parameter study
together with all the available observational data including spectra
on NGC 7252 to trace back the SFH of this Sc - Sc merger remnant
(Fritze -- v. A. \& Gerhard 1994b). We find evidence for a strong
global starburst that started $\sim 1.3$ Gyr ago and had a duration of
$\sim 10^8$ yr, and we predict that -- if present SF should stop --
NGC 7252 would reach typical elliptical galaxy colors within $1.5-3.5$
Gyr. Instead, however, if the molecular gas disk will be transformed
into stars, NGC 7252 may end up as a disky E or even as an S0 galaxy.
The nuclear spectrum of NGC 7252 shows H$_{\beta}$ emission at the
bottom of the deep absorption, HST reveals a strong bluing
($\Delta$(V-I)$\sim 0.5$ mag) within the central 200 pc (Whitmore
\etal 1993) and IUE observations indicate ongoing SF at a rate of
$\sim 4$ M$_{\odot}$yr$^{-1}$ (Fritze -- v. A. \& Schweizer, {\it in
prep.}). Our best fit model has a present gas restorage rate from
dying (burst) stars of $\gta 2 \msun\yr^{-1}$ which was much higher in
the recent past.  Moreover, the HI falling back from the tails and not
observed within the body of NGC 7252 might, at least partly, be
transformed into molecular gas replenishing the reservoir for SF.
{\bf In summary}, it looks as if we were whitnessing here the
evolution of a strong global starburst $\sim 1$ Gyr ago into a weak
central burst still active today and the building up of an appreciable
stellar disk in an elliptical-like merger remnant. \hb On the basis of
the progenitor spirals' ISM properties we predict metallicities of
stars and star clusters formed in the burst:
Z$\,\gta\,\third\,\cdot\,$Z$_{\odot}~~\leftrightarrow ~~$
[Fe/H]$\,\gta -0.8$. HST detection of a substantial population of
Young Star Clusters ({\bf YSC}) (Whitmore \etal 1993) and spectroscopy
of the two brightest of them (Schweizer \& Seitzer 1993) confirm the
high SF efficiency over a large volume of the galaxy ($R \lta 14$ kpc)
implied by our high burst strength, our metallicity prediction, the
starburst age, and the idea of a central afterburning (Fritze --
v. A. \& Burkert 1995). With ages $\lta 1.3$ Gyr, most of these YSCs
should be globular clusters rather than open ones most of which would
already have dispersed.

\section{NGC 4038/39: Formation of Young Star Clusters}
NGC 4038/39 (= Arp 244 = ``The Antennae'') is the youngest interacting
galaxy pair in Toomre \& Toomre's list, an ongoing merger with a
starburst that began $\sim 2 \cdot 10^8$ yr ago. I have chosen it to
discuss the HST detection of $\sim 700$ YSCs (Whitmore \& Schweizer
1995), typically a dozen of which, themselves, are clustered within
giant HII regions, relics of the supergiant molecular cloud they were
born from. This already hints at a molecular cloud structure different
from that in the Milky Way ($\rightarrow$ Sect. 5).

YSCs are seen in many other interacting galaxies/mergers, too: NGC
3597 (Lutz 1991), NGC 1275 (Holtzman \etal 1992), NGC 1140 (Hunter
\etal 1994), NGC 1705 and M82 (O'Connell \etal 1994, 95), He2--10
(Conti \& Vacca 1994), Cartwheel (Borne 1995), and several examples in
Meurer \etal (1996).

A sharp controversy is centered on the question if these YSC are young
Globular Clusters ({\bf GC}) or rather open
clusters/OB-associations. Discriminating properties are the effective
radii R$_{{\rm eff}}$ and the Luminosity Function ({\bf LF}) of the
YSCs (van den Bergh 1995).  Indeed, observations give a wide range of
effective radii, however, Meurer (1995) argues that because of
crowding of the YSCs on a bright and variable background these
effective radii are probably largely overestimated. They estimate the
distance out to which effective radii of YSCs can reliably be measured
to be $\sim 9$ Mpc and find that for all 3 galaxies close enough with
YSCs detected the mean R$_{{\rm eff}}$ of YSCs are indeed well within
the range observed for Milky Way GCs.  We model the spectral and
photometric evolution of star clusters for different initial
compositions ($10^{-4} \leq$ Z $\leq 2 \cdot$ Z$_{\odot}$) and find a
strong effect of the metallicity on the color evolution, already at
early stages. Knowing the metallicity of a YSC from spectroscopy (or
estimating it from the ISM properties at its birth) then allows for
quite precise age dating and for the prediction of its future
luminosity evolution (Fritze -- v. A. \& Burkert 1995). The young mean
age of $\sim 2 \cdot 10^8$ yr of the YSCs in the Antennae, for which
lack of spectroscopy we can only assume the same metallicity as that
of the YSCs in NGC 7252, makes it evident that open clusters may well
be coexistent with young GCs. We therefore divide the YSCs into
subsamples with large and small R$_{{\rm eff}}$, cutting somewhat
arbitrarily at R$_{{\rm eff}}\,=\,10\,$pc because of the above
mentioned overestimation. Assuming a common age of $2\,\cdot\,10^8\,$
yr for the YSCs we then evolve the two LFs over a Hubble time and find
that they look significantly different. While the LF for clusters with
R$_{{\rm eff}}\,>\,10\,$pc looks exponential like the LF of Galactic
open clusters, the LF for YSCs with R$_{{\rm eff}}\,<\,10\,$pc more
resembles the Gauss-shaped LF of Galactic GCs with the maximum of the
distribution at approximately the correct $M_V \sim -7.4$ (Secker
1992). However, it features a strong overpopulation in the faint bins
(Fritze -- v. A. 1995).  It is clear that for an ongoing starburst as
in the Antennae the age spread among the YSCs should not be
neglected. This age spread causes the faint clusters in the present LF
to be older on average than the bright ones and, consequently, to fade
less over the rest of the Hubble time, moving them to brighter bins in
the LF (Fritze -- v. A. \& Burkert, {\it in prep.}).  Dymanical
effects like evaporation through internal stellar mass loss as well as
loss of stars and destruction in the inhomogeneous tidal field of the
interacting galaxy pair are very difficult to quantify but might
preferentially destroy low mass (=low luminosity) clusters.

{\bf Summarising}, I see substantial evidence for the possibility that
in gas-rich mergers a significant population of GCs can be
formed. These second generation GCs have higher metallicities:
[Fe/H]$\gta -0.8$ for mergers of Sb$\dots$Sd galaxies that happened
not more than 5 Gyr ago. Detection of a bimodal metallicity
distribution in a GC system can prove the Sp -- Sp merger origin of
the parent E/S0 galaxy long after other suspicious features (tails,
ripples, plumes, shells, ...) have disappeared.  And if, indeed, GCs
form during mergers in numbers comparable to those typical for spirals
(Zepf \& Ashman 1993), CA 4 is cancelled.

Whatever the nature and future fate of these YSCs, they do trace the
dynamical evolution of the starburst through the fact that -- on
average -- they are observed at those galactocentric distances where
they were born. Because of easier background subtraction they are much
better suited in this respect than the integrated stellar population.
A first analysis of radial trends in mean values and rms-scatter of
colors, luminosities, and R$_{{\rm eff}}$ are consistent with the
scenario of a global starburst contracting with time, like in NGC 7252
(Fritze -- v. A. \& Burkert, {\it in prep.}).  However, spectroscopy
of the YSCs is required to determine precise ages and metallicities
before any firm conclusions can be reached.  The very youngest star
clusters might even allow to explore the upper IMF.

\section{Formation of Dwarf Galaxies in Tidal Tails}
N-body as well as TREE-SPH models have shown that condensations along
tidal tails can form and become self-gravitating with masses typical
of dwarf galaxies (Barnes \& Hernquist 1992b, Elmegreen \etal
1993). Tidal tails contain material from the spiral disks, so these
dwarf galaxies will not have DM halos and it is not evident that they
will be able to survive the starburst which forms them.  The two
condensations in Arp 105 studied by Duc \& Mirabel (1994) are found to
resemble a BCDG and an Im galaxy, respectively. They show [O/H] values
typical of outer parts of spirals and M/L $\sim 1$ inside the optical
radius. With luminosities of Im's or BCDGs, these objects resemble
{\it faint blue galaxies}. In the past, mergers were more frequent
($\rightarrow$ Sect. 8) and galaxies contained more gas, so it may
well be conjectured that fading (and dynamically transient?)  tidal
dwarf galaxies may have to do with the fading of a large dwarf galaxy
population invoked to explain the faint galaxy counts, even for
$\Omega =1$ (e.g. Phillipps \& Driver 1995).  The gas-rich merging
Superantennae (=IRAS 19254-7245) shows a whole chain of dwarf-like
objects along its gigantic tails (Mirabel \etal 1991). Clearly, more
observations are needed to test this hypothesis, in particular
spectroscopy to disentangle age and metallicity before the fading can
be calculated as well as kinematic information and surface brightness
profiles to enable dynamical modelling. A problem, if confirmed, might
be the weak clustering of the faint blue galaxies.

\section{Arp 220: Molecular Cloud Structure in Gas-Rich Mergers}
I choose my $3^{{\rm rd}}$ example to be Arp 220, an advanced merger
and IR-UL galaxy to discuss the Molecular Cloud ({\bf MC}) structure
in massive gas-rich interacting galaxies. Arp 220 has a total L$_{{\rm
IR}} \sim 1.6 \cdot 10^{12} \lsun$ with $\twothirds$ L$_{{\rm CO}}$
coming from a nuclear disk of r$\,\sim 300$ pc and $\third$ L$_{{\rm
CO}}$ from an extended ($\sim 5$ kpc) component (Scoville \etal 1991).
To determine the MC structure in Arp 220 and many other IR-UL galaxies
advantage is drawn from the fact that there are submm lines that trace
molecular gas at very different densities, e.g. CO($1\rightarrow0$) at
n$\,\sim 500$ cm$^{-3}$, HCN($1\rightarrow0$) at n$\,\sim 10^4$
cm$^{-3}$, CS($2\rightarrow1,~3\rightarrow2$) at n$\,\sim 10^5$
cm$^{-3}$. HCN observations of Arp 220 (and other IR-ULs) reveal
central gas densities $\rho_0 \lta 500\,\msun$ pc$^{-3}$ comparable to
central stellar densities in ellipticals (eg. Solomon \etal 1992). The
molecular gas dominates the dynamical mass in the centers of these
galaxies. On a scale of $\lta 1$ kpc, the core of Arp 220 looks like
{\bf one supergiant MC core} and comparison of L$_{{\rm CO}}$ with
L$_{{\rm CS}}$ shows that it contains comparable amounts of ultradense
and low-density gas ($\sim 10^{10} \msun$), while a giant MC in the
Milky Way, the LMC, or a BCDG typically only contains few percent of
its total mass at densities visible in CS, i.e. high enough to be
transformed into stars. All kinds of SF indicators better correlate
with L$_{{\rm CS}}$ or L$_{{\rm HCN}}$ than with L$_{{\rm CO}}$.  SF
efficiencies $\eta$, i.e. the fraction of the gas mass that is
transformed into stars, are also different by up to two orders of
magnitude between IR-UL galaxies/mergers ($\eta \sim 0.1 \dots 0.5$,
Fritze -- v. A. \& Gerhard 1994b) and spirals/Im's/BCDGs ($\eta \sim
0.001 \dots 0.01$).  Despite some uncertainty in the L$_{{\rm
CO}}$--to--M(H$_2$) conversion factor it is clear that the MC
structure is drastically different in IR-UL galaxies and other
starforming environments.  This may rise some doubts as to the
universality of the SF process itself.  The cloud -- cloud collision
scheme for SF seems to break down in the dense core of Arp 220.  Is
there a violent mode of SF realised in massive gas-rich mergers and
possibly in an initial collapse as opposed to a peaceful mode now
whitnessed in spirals, irregulars and even in the comparatively tiny
starbursts in BCDGs?  To further investigate this question, the study
of starburst parameters should be extended to the NIR and applied to
samples of interacting and IR-UL galaxies.

\section{The Interaction -- Activity Connection}
Optical imaging shows that among the IR luminous galaxies the fraction
of interacting systems increases with IR-luminosity from $\sim 25 \%$
for L$_{{\rm CO}} < 10^{11} \lsun$ to 50 -- 70\% for $10^{11} \leq$
L$_{{\rm CO}}/\lsun \leq 10^{12}$, and to $\lta 100 \%$ for L$_{{\rm
CO}} > 10^{12} \lsun$ (Sanders \etal 1988, Melnick \& Mirabel 1990,
Leech \etal 1994, Clements \etal 1995). While in many cases all of the
FIR emission is explained by strong nuclear starbursts with SFRs of
order $100~ \msun $yr$^{-1}$ some hyperluminous IR galaxies may also
contain a dust-enshrouded AGN (Hines \etal 1995, Goldader \etal 1995).
More than 20 years ago, Searle \etal (1973) and Larson \& Tinsley
(1978) already recognised that the morphologically peculiar galaxies
from Arp's catalogue on average have bluer $UBV$ colors than normal
galaxies and they interpreted these blue colors in terms of enhanced
SF activity.

A connection between non-thermal activity and interactions is also
found for Seyfert galaxies, many of which are in small groups with a
low velocity dispersion and a high spiral fraction, i.e. in an
environment most favorable for strong interactions involving gas-rich
galaxies (Dahari 1984,...).  Some Seyfert galaxies also show tidal
tails or close companions (Fricke \& Kollatschny 1989, Monaco \etal
1994).  Circumnuclear starburst and postburst features are detected
around an increasing number of Seyfert galaxies (e.g.  Genzel \etal
1995, Oliva \etal 1995) with a possible trend of increasing starburst
luminosity with increasing AGN luminosity (Goerdt \etal 1993).
Hutchings \etal (1989) show that $\sim 70 \%$ of the nearby QSOs have
close companions. QSO host galaxies often show distorted
morphpologies, tidal features and/or multiple nuclei (e.g. Stockton \&
Ridgeway 1991). Dickinson (1995) reports high redshift QSOs to be
located near the centers of galaxy clusters in formation. High
redshift radio galaxies are frequently seen to feature dust-lanes,
tidal tails and bridges, shells and double nuclei, all of which may
point to ongoing or recent interactions. \hb Thus, it seems clear that
{\bf interactions can trigger both thermal (=starburst) and
non-thermal (=AGN) activity.} However, there is some strange
interaction -- activity dichotomy (courtesy K. Borne) in the sense
that many active galaxies show signs of interactions while not all
interacting galaxies do show signs of activity.  A solution to this
dichotomy might be that evidence of tidal damage may be long-lived
(few Gyr) whereas any form of activity is short-lived ($\sim 10^8$
yr). This hypothesis is the subject of many ongoing observational and
theoretical investigations. \hb {\bf To summarise:} interactions
involving at least one gas-rich galaxy may lead to activity which, in
turn, may range from global starbursts (eg. NGC 7252) to nuclear ones
(eg. NGC 7714 and many IR-UL galaxies) and to AGNs (QSOs, Seyfert, and
hyperluminous IR galaxies). The obvious question is, of course: Is
there a transition between these 3 features -- either in time or in
parameter space (e.g. gas content, velocity dispersion, geometry of
encounter, bulge:disk:halo-ratio, $\dots$)?  A time evolution from a
global towards a nuclear starburst seems indicated in case of NGC 7252
and ``The Antennae'', a transition from starburst to AGN activity by
observations of circumnuclear bursts/postbursts in Seyfert galaxies,
comparable luminosity contributions from starbursts and
dust-enshrouded QSOs in case of hyperluminous IR galaxies (Vielleux
\etal 1995) and from observations of comparable CO-luminosities in
IR-ULs and QSOs (Barvainis \etal 1994, Lonsdale \etal 1995). \hb A
severe problem for numerical models is the enormous dynamical range
extending from $\sim 100$ kpc down to $\leq 10$ pc.  A first attempt
to bridge all of this range are Bekki's (1995) TREE SPH models of
merging spirals including a simple SF prescription. They show that the
orbiting cores {\it stir up} the gas clouds precipitating SF and that
dissipative cloud-cloud collisions are able to drive $\sim 10^8~
\msun$ of gas into $r \leq 10$ pc shortly after core merging. In this
model, a {\bf starburst} dominates L$_{{\rm bol}}$ {\bf before core
merging} while an accreting {\bf AGN} takes over {\bf thereafter} with
retrograde encounters being most efficient in fuelling the
nucleus. The low percentage of 8 out of 243 Seyfert galaxies from
V\'eron \& V\'eron (1989) showing double nuclei (Kollatschny \& Fricke
1989) is compatible with this scenario of AGN activity occuring in
late stages/after completion of a merger.

\section{Merger Remnants among Present Day Normal Galaxies}
It came as a big surprise when a second nucleus 2 pc away from and
brighter than the one at the center of the isophotes was detected in
Andromeda (Lauer \etal 1993, Gerssen \etal 1995). Shells and warps
reminiscent of past interactions are seen in several spiral galaxies
(Schweizer \& Seitzer 1988, ...). For the Milky Way, the 2-component
GC system, the retrograde halo stars, the Magellanic stream, as well
as the thick disk may be indicating past interactions. While Toth \&
Ostriker (1992) used the existence of a thin disk to limit the rate of
satellite accretion, Athanassoula \& Bosma (1987) shows that massive
halos can stabilise galactic disks considerably so that they can
``digest'' without damage up to $10 \%$ of their own mass, provided
the intruder is not too compact. Hernquist's models as well as
observations of a molecular gas disk in NGC 7252 show that disks can
partly be rebuilt after tidal destruction.  The prolongued backfall of
gas from tidal tails supports this process.  Mergers of galaxies with
a mass ratio of 3:1 even do leave disk-like remnants in many cases
(Barnes 1995).  Thus, while among spirals, there is a lot of evidence
for past interactions, no statistical analysis of the fraction of
nearby spirals having experienced an interaction of a given strength
is available. \hb For E/S0 galaxies, Schweizer \etal (1990) define a
fine structure parameter $\Sigma$ meant to somehow quantify the
deviation from an ``ordinary'' appearance and show that the amount of
fine structure in a galaxy correlates with its EW(H$_{\beta}$) and
optical colors in the sense that much fine structure goes with a
younger mean age of the stellar population (Schweizer \& Seitzer
1992). They give a mean ``heuristic merger age'' (= time since last
major merger) of $5-10$ Gyr for the E/S0 galaxies from their sample.
Peculiar core kinematics as well as disky isophotes are observed in
$\sim 30 \%$ of Virgo's ellipticals (Bender 1995, Skorza 1995).
X-coronae are interpreted as coming from hot gas expelled from the
galaxy by a starburst-driven superwind (e.g. Fabbiano \etal 1989).
The levelling-off above luminosities of $10^{11.5} \lsun$ of Mg$_2$
gradients that increase with luminosity from L $\sim 10^{10}
\rightarrow 10^{11.5} \lsun$ is taken as evidence for dissipative
effects coming into play in the most luminous ellipticals.  Hiu \etal
(1995) report PNe and GCs to significantly rotate around major and
minor axes in the outer halo of the dust-lane galaxy NGC 5128 (= Cen
A). This outer halo rotation is in agreement with Hernquist's (1993)
prediction for merger remnants.  Intriguing coincidences between core
mass, core radius, and core density of ellipticals and CO-mass,
CO-radius, and gas density in IR-UL galaxies (Doyon \etal 1994) may
(but need not) point to a Sp -- Sp merger origin of many ellipticals.
The tightness of the Fundamental Plane ({\bf FP}) is often used as an
argument against the merger origin of ellipticals.  However, while
dissipationless mergers were shown to conserve the FP (Capelato \etal
1995), the analysis of E -- Sp interactions yields a ``FP of changes''
that is only inclined by $15 \deg$ with respect to the observed one
(Levine 1995).  Lake \& Dressler's (1986) observations of 16 recent
mergers from Arp's atlas show them all to lie perfectly within the FP
of normal ellipticals. \hb {\bf To conclude:} the FP seems to be the
result of any violent relaxation process (involving quantities at $r
\leq r_e$ where relaxation is fast) and it is not suitable to test
for/against a merger origin of ellipticals.  We have, however, already
shown in Sect. 3 that the metallicity distribution of a galaxy's GC
system, observable at least up to Virgo cluster distances, does allow
to reveal the origin of E/S0 galaxies (see also Zepf \& Ashman 1993).
In case of a classical monolithic collapse scenario a narrow [Fe/H]
distribution is expected for the GC system and tentatively reported by
Kissler -- Patig \etal (1995) for several Fornax dEs. The bimodal
metallicity distribution with a secondary maximum around [Fe/H] $\geq
-0.8$ predicted for Sp -- Sp merger remnants is observed e.g.  in NGC
4472, NGC 5128 (Zepf \& Ashman 1993), NGC 3923 (Zepf \etal 1995), and
NGC 3256 (Zepf 1995).  For the GCs of cD galaxies with their complex
and prolongued merging/accretion history a broad or multiply peaked
[Fe/H] distribution is expected as indeed observed e.g. in NGC 1399
(Lee \& Geisler 1993).

\section{Merger Rates}
From statistics of nearby galaxy pairs Toomre (1977) estimates a
merger rate of $\sim 0.1$ t$_{{\rm H}}^{-1}$, the recent analysis of
Keel \& Wu (1995) gives $\sim 0.33$ t$_{{\rm H}}^{-1}$, t$_{{\rm H}}$
being the Hubble time.  In the past, merger rates were undoubtedly
higher. \hb Theoretical models based on Press--Schechter formalism
with a decoupling from the merging hierarchy predict a redshift
evolution of the merger rate $\sim (1+{\rm z})^{4.5 \Omega ^{0.42}}$
(Carlberg 1990), the CDM halo merging model of Lacey \& Cole (1993)
also gives a strong increase with redshift with the exact steepness
depending on the mass scale involved. \hb Observations give redshift
evolutions of the merger rate as $\sim (1+{\rm z})^{2.5}$ (Toomre 1977
for $\Omega=1$), $\sim (1+{\rm z})^{4 \pm 2.5}$ (Zepf \& Koo 1989 from
POSS plates and $\sim 1000$ galaxy redshifts), $\sim (1+{\rm z})^{2.5
\pm 0.5}$ (Burkey \etal 1994 from HST observations of 146 galaxies to
$I=26$), $\sim (1+{\rm z})^{3.4 \pm 1.0}$ (Carlberg \etal 1994, 410
galaxies from the CFHT redshift survey to $V=22.5$), and $\sim (1+{\rm
z})^{4.0 \pm 1.5}$ (Yee \& Ellingson 1995 for 100 galaxy redshifts
from the QCRS). \hb So, despite some quantitative differences, both
theory and observations agree in a strong increase of the merger rate
to higher redshifts. Significant merger rates in the past are expected
to reshape the luminosity function of galaxies, to bring along galaxy
number density evolution and to change the Hubble type mix with
redshift, as well as to add some extra luminosity during
merging-triggered starbursts. Some of these effects have been included
one-by-one via simple parametrisations into models for the
interpretation of galaxy counts ($\rightarrow$ eg. Guiderdoni \& Rocca
-- Volmerange 1991, Carlberg \& Charlot 1992), a consistent treatment
of all effects together, however, has not been attempted until
now. Necessary prerequisits for this kind of modelling are to
determine local values for burst strengths and durations which we are
currently attempting using Liu \& Kennicutt's (1995) sample of
interacting galaxies. Still, these values will only apply for
low-redshift galaxies, and need not be the same in the early universe
where galaxies were much more gas-rich. Furthermore, a better
understanding of the longterm evolution of merger remnants (SFRs,
rebuilding of disks, $\dots$) as well as of the whole range of
interaction events (mass ratios, relative velocities, orbital
parameters, etc.) and their effects on the galaxies involved is
required. \hb At any given redshift, the merger rate is expected to
depend on environment. Structure formation scenarii (e.g. Kauffmann \&
White 1993) predict that in low density environments merging should
occur later than in high galaxy density regions. This prediction is
consistent with de$\,$Carvalho \& Djorgovski's (1992) finding that at
fixed luminosity field galaxies are bluer, have lower Mg$_2$ and
higher surface brightness than cluster ellipticals, which can be
understood in terms of field E's being younger than cluster E's. \hb
How spectacular a merger will look like also depends on environment:
Barnes (1992) shows that, within a group or cluster, the tidal forces
of the neighbors tend to shred the tidal tails. If only one member of
a galaxy pair is gas-rich, the result of a merger might be much less
spectacular, both morphologically and spectroscopically, than the
famous examples of two gas-rich spirals. In a galaxy cluster at low
redshift, the slow encounters favorable for galaxy mergers are very
rare, instead, the rapid fly-by's may also transform Hubble types and
cause starbursts by destabilising gas disks of infalling spirals.
These effects of {\it harassment} are studied e.g.  by Moore (1995).

All of these environmental aspects as well as the possibility of
multiple mergers (Weil \& Hernquist 1996) are important for the
discussion of the nature of Butcher -- Oemler galaxies in distant
galaxy clusters ($\rightarrow$ Poggianti, Barger {\it this volume}).

\section{Outlook}
Only briefly touching some specific aspects of {\it Interacting
Galaxies} I did not at all mention many interesting wavelength ranges
as e.g. X-, UV-, NIR- and IRAS observations. I tried to show that
dynamical models are in pretty good shape and continuously refined,
the major obstacle at present being our poor understanding of the SF
process. I hope that interacting galaxies, when carefully analysed
with spectrophotometric, dynamical, and chemical models -- making use
of the young star cluster population -- may help improve our knowledge
even about star and star cluster formation.  We need a solid basis in
observations and interpreting models at low redshift. Models
attempting to understand high redshift galaxies need to consistently
account for all effects related to merging and the possible triggering
of starbursts.  Galaxy clusters at intermediate and high redshift may
provide important clues as to galaxy population, merger rates, etc.,
provided environmental effects as well as the evolutionary states of
the clusters themselves are properly accounted for.  A particularly
promising approach seems to attempt a combination of galaxy formation
scenarii with spectrophotometric and chemical evolution models.

\acknowledgments I am deeply indebted to Kirk Borne, Steve Zepf, Brad
Whitmore, and Francois Schweizer for many inspiring discussions, and
to Francois Schweizer for valuable comments on an earlier draft of the
manuscript.  This work has been supported by a Habilitationsstipendium
of the Deutsche Forschungsgemeinschaft under Fr. 316/2-1.

% That's all, folks.
%
% The technique of segregating major semantic components of the document
% within "environments" is a very good one, but you as an author have to
% come up with a way of making sure each \begin{whatzit} has a corresponding
% \end{whatzit}.  If you miss one, LaTeX will probably complain a great
% deal during the composition of the document.  Occasionally, you get away
% with it right up to the \end{document}, in which case, you will see
% "\begin{whatzit} ended by \end{document}".

\end{document}